# Robust Solid State Quantum System Operating at 800 K

Mehran Kianinia, Sherif Abdulkader Tawfik, Blake Regan, Toan Trong Tran, Michael J. Ford, Igor Aharonovich, [*] and Milos Toth[^]

[†] School of Mathematical and Physical Sciences, University of Technology Sydney, Ultimo, NSW, 2007, Australia.

**ABSTRACT**: Realization of Quantum information and communications technologies requires robust, stable solid state single photon sources. However, most existing sources cease to function above cryogenic or room temperature due to thermal ionization or strong phonon coupling which impede their emissive and quantum properties. Here we present an efficient single photon source based on a defect in a van der Waals crystal that is optically stable and operates at elevated temperatures of up to 800 K. The quantum nature of the source and the photon purity are maintained upon heating to 800 K and cooling back to room temperature. Our report of a robust high temperature solid state single photon source constitutes a significant step towards practical, integrated quantum technologies for real-world environments.

## INTRODUCTION

Modern light-based technologies require sophisticated materials growth and device engineering techniques to achieve optimum performance under ambient conditions. Optically active defects in solids are a vital part of numerous such technologies spanning light emitting diodes (LEDs)[1, 2], lasers[3], sensors[4, 5] and communications[6]. An emerging application of localized individual defects is their use as single photon sources (SPSs) in integrated nanophotonics and quantum information processing schemes that utilize individual photons as information carriers[7, 8, 9, 10]. For such applications to become a reality there is an urgent need for robust, on-demand SPSs. Furthermore, these sources should be optically active in harsh environments and at elevated temperatures that are typical of high density integrated circuits.

Hence, over the last decade, research into solid state SPSs – quantum dots (QDs) and color centers – has accelerated dramatically. Quantum dots mostly operate only at cryogenic temperatures because of low carrier binding energies and thermal ionization at elevated temperatures[11, 12]. On the other hand, color centers in solids (defects and impurities, also known as artificial atoms) often have deep electronic states that enable SPS operation near room temperature (RT)[13, 14, 15]. However, both QDs and color centers typically suffer from efficient coupling between electrons and lattice phonons that results in an overall reduction of brightness at higher temperatures.

In this work, we report a solid state SPS that operates at temperatures as high as 800 K. The SPS maintains its quantum nature during a high temperature thermal cycle – that is, the photon purity (defined as the fraction of single photon emission events) does not decrease upon heating, and the SPS optical properties are preserved after heating up to 800 K and cooling back to room temperature, as would be required for practical on-chip devices. The SPS is a deep trap defect in layered hexagonal boron nitride (hBN) – a van der Waals crystal with a wide indirect bandgap of ~ 6 eV[16] and favorable thermal, chemical and mechanical properties[17, 18]. The recently-discovered SPSs in hBN are extremely bright, chemically stable, fully polarized, and exhibit a broad range of emission colors[19, 20, 21, 22, 23, 24, 25]. Given the two dimensional (2D), layered nature of hBN and the extreme thermal stability of the SPSs demonstrated here, fabrication of hybrid photonic systems is a promising potential path to high density integrated circuits designed to withstand heating that occurs during operation at RT.

## MATERIALS AND METHODS

A schematic illustration of hBN on a hot substrate is shown in figure 1a. The sample consists of a few-layer flake of atomically thin hBN bonded by van der Waals forces. The hBN flakes (Graphene Supermarket) were dropcast onto a silicon wafer and annealed in a tube furnace at 850 °C for 30 minutes under 1 Torr of Argon to activate the SPSs[19]. To perform the heating experiments, we use a purpose-built vacuum chamber with a thin quartz window and a long working distance, high numerical aperture (0.7) objective, shown in figure 1b. The samples were mounted on a pyrolitic boron nitride heater, and the temperature was measured using a thermocouple fixed to the silicon substrate. A turbo molecular pump was used to achieve a base pressure of $\sim 1\times 10^{-6}$ mBar, and a 532 nm laser was used as the excitation source in the confocal photoluminescence setup. For all measurements except time-resolved photoluminescence, the laser power was fixed at 3 mW (measured before the window of the chamber). The collected light was filtered using a 532 nm dichroic mirror and an additional long pass filter. The collected light was sent either to a spectrometer (Acton Spectra ProTM, Princeton Instrument Inc.) equipped with a 300 lines/mm grating and a charged-coupled device (CCD) detector with a resolution of 0.14 nm or to a Hanbury Brown and Twiss set-up for single photon correlation measurement (figure 1c).

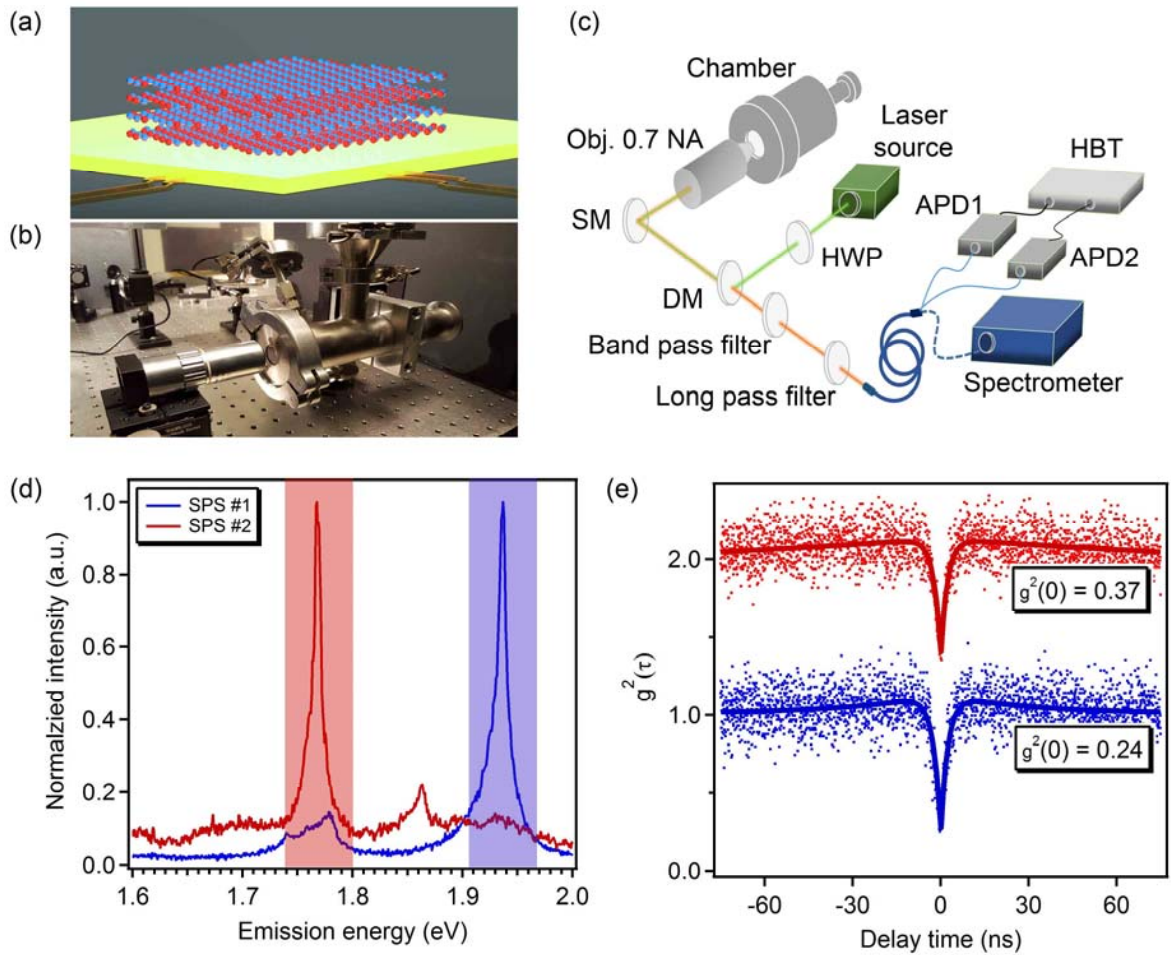

Figure 1. (a) Schematic illustration of a hBN crystal hosting a single photon source atop of a heated substrate. (b) The purpose-built vacuum chamber setup used in this study. (c) Hanbury Brown and Twiss set-up equipped with vacuum chamber used in this study. Band pass filter according to ZPL energy of the SPS was used only for conscience measurements. (d) Room temperature PL spectra acquired from two representative SPSs in hBN. The shaded areas represent the filtered regions used to record the correlation measurements. (e) Antibunching behavior of the sources demonstrated by autocorrelation functions with $g^2(0)$ values of 0.24 and 0.37 for SPS #1 (blue) and SPS #2 (red), respectively, confirming the quantum nature of the emissions. The $g^2(\tau)$ from SPS #2 was offset for clarity. The data were fitted with a three level system (solid curves).

The emitted light was coupled to an optical fiber and split to two paths into avalanche photodiodes in order to perform correlation measurements using a time correlated single photon counting module (Picoharp300TM, PicoQuantTM). Lifetime measurements were done with a 510 nm pulsed laser excitation source at 30 μw power with a 100 ps pulse width and 10 MHz repetition rate.

## RESULTS AND DISCUSSION

Two SPSs in different hBN crystals were studied in this work. Figure 1c shows RT photoluminescence (PL) spectra recorded from the SPSs. The spectrum from SPS #1 consists of two peaks at ~1.94 eV and ~1.78 eV attributed to the zero phonon line (ZPL) and the phonon side band (PSB), respectively[19]. SPS #2 has a sharper ZPL centered at ~1.75 eV, and a negligible PSB (the emissions beyond ~1.8 eV are background). These particular two emitters were selected for the present study because their spectra are representative of the emission spectrum diversity that has been observed previously in hBN[19] (in terms of ZPL width and position, and PSB intensity).

After we identified reliable SPSs, we proceeded to the heating experiment. The temperature-dependence of each SPS was characterized by measuring its photophysical properties at 100 K increments during a 300 K – 800 K – 300 K thermal cycle. Figure 2a, b and figure 2c, d show autocorrelation functions and normalized PL spectra obtained from SPS #1 (blue data set) and SPS #2 (red data set) obtained during the heating phase of the thermal cycle, respectively.

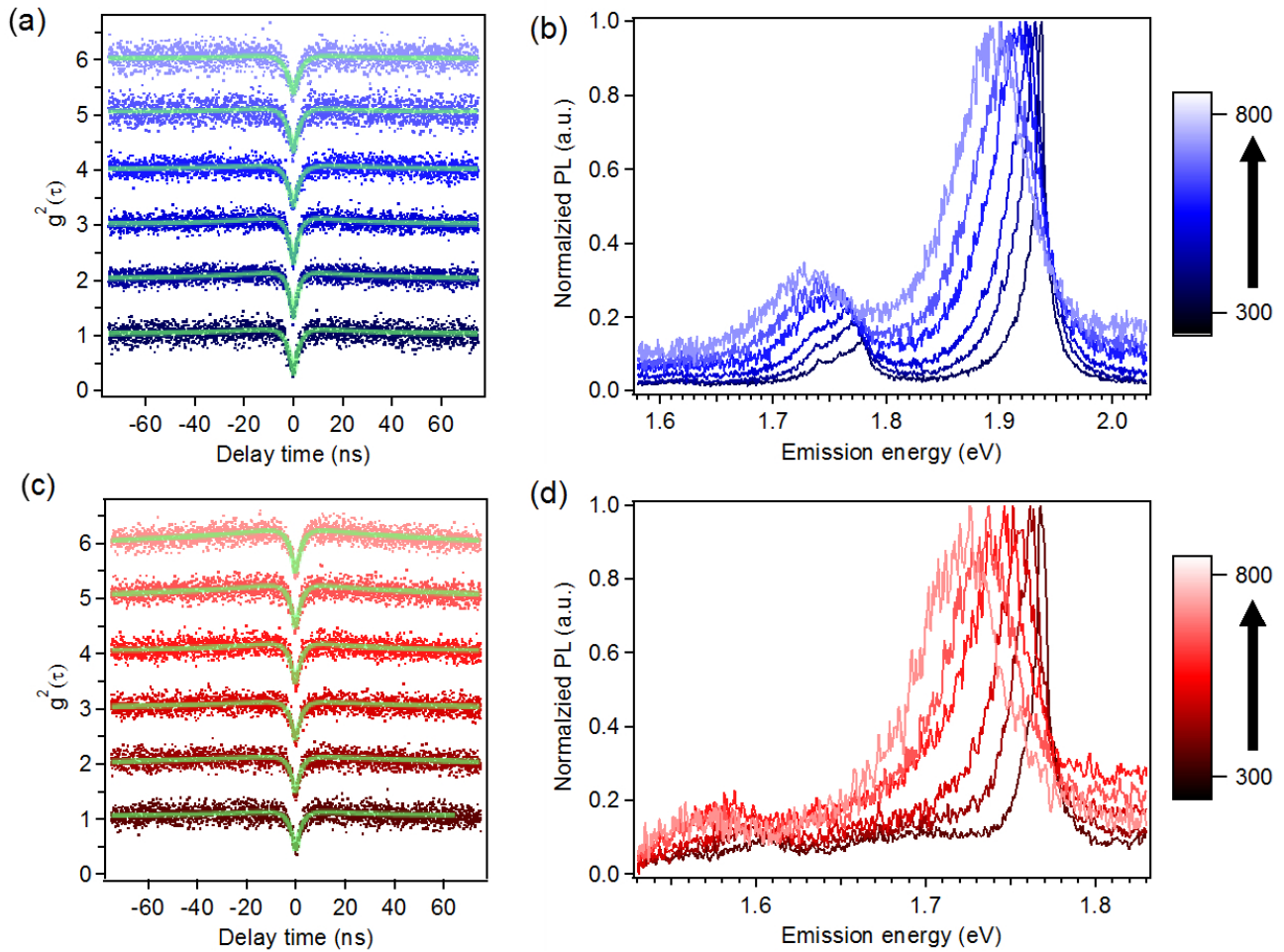

Figure 2. Temperature dependence of the quantum emission and photoluminescence of SPSs in hBN during heating. (a) Autocorrelation curve and (b) normalized PL spectrum measured from SPS #1 at 100 K intervals between 300 K and 800 K. (c-d) Equivalent data from SPS #2. The autocorrelation functions in (a) and (c) were offset for clarity. Green solid curves are fits obtained using a three level model.

Equivalent data acquired during cooling are shown in figure 3. The single photon emission persists all the way to 800 K, and the $g^2(0)$ values are independent of temperature, as is discussed below. The data show unambiguously that both sources are stable and can operate during and after exposure to temperatures as high as 800 K. The spectra red shift and broaden with increasing temperature, and the changes are reversible upon cooling back to RT (figure 3), demonstrating that the SPSs are not compromised by the thermal cycle. At 800 K, the ZPLs are centered at ~ 1.90 eV and 1.72 eV, and have a FWHM of ~ 80 meV and 70 meV for SPS #1 and SPS #2, respectively.

To characterize the photon emission statistics and show the quantum nature of the SPSs, second order autocorrelation functions, $g^2(\tau) = \frac{<I(t)I(t+\tau)>}{<I(t)>^2}$, were acquired using a Hanbury Brown and Twiss (HBT) interferometer. The shaded rectangles in figure 1d represent the filters used to collect the photons in the autocorrelation measurement. Figure 1e shows the $g^2(\tau)$ curves recorded from each source at RT and fitted using a three level model. The antibunching dip below 0.5 at zero delay time ($\tau = 0$) constitutes proof for the quantum nature of a photon source. The raw (uncorrected) values of $g^2(0)$ are 0.24 and 0.37 for SPS #1 and SPS #2, respectively. The background corrected autocorrelation data suggest that the deviation from $g^2(0) = 0$ is due to residual background. The autocorrelation data were background corrected using the procedure described in reference[26]. Signal to background ratios (S/B) were calculated based on the counts acquired when the excitation laser was focused on the source and when it was focused on the adjacent silicon substrate. The signal and background were measured at each temperature and the corrected coincidence ($g_c^2$) was obtained using the equation, $g_c^2(\tau) = \frac{g^2(\tau)-(1-\rho^2)}{\rho^2}$, where $\rho$ is the signal-to-noise ratio S/(S+B) and $g^2(\tau)$ is the uncorrected photo-coincidence. The excited state lifetime was measured using a pulsed laser, and remained almost the same, around 3.5 ns, at all temperatures for both sources (figure 4a,c). In addition, both sources were stable and no blinking was observed at any temperature during heating to 800 K or cooling back to RT (figure 4b,d).

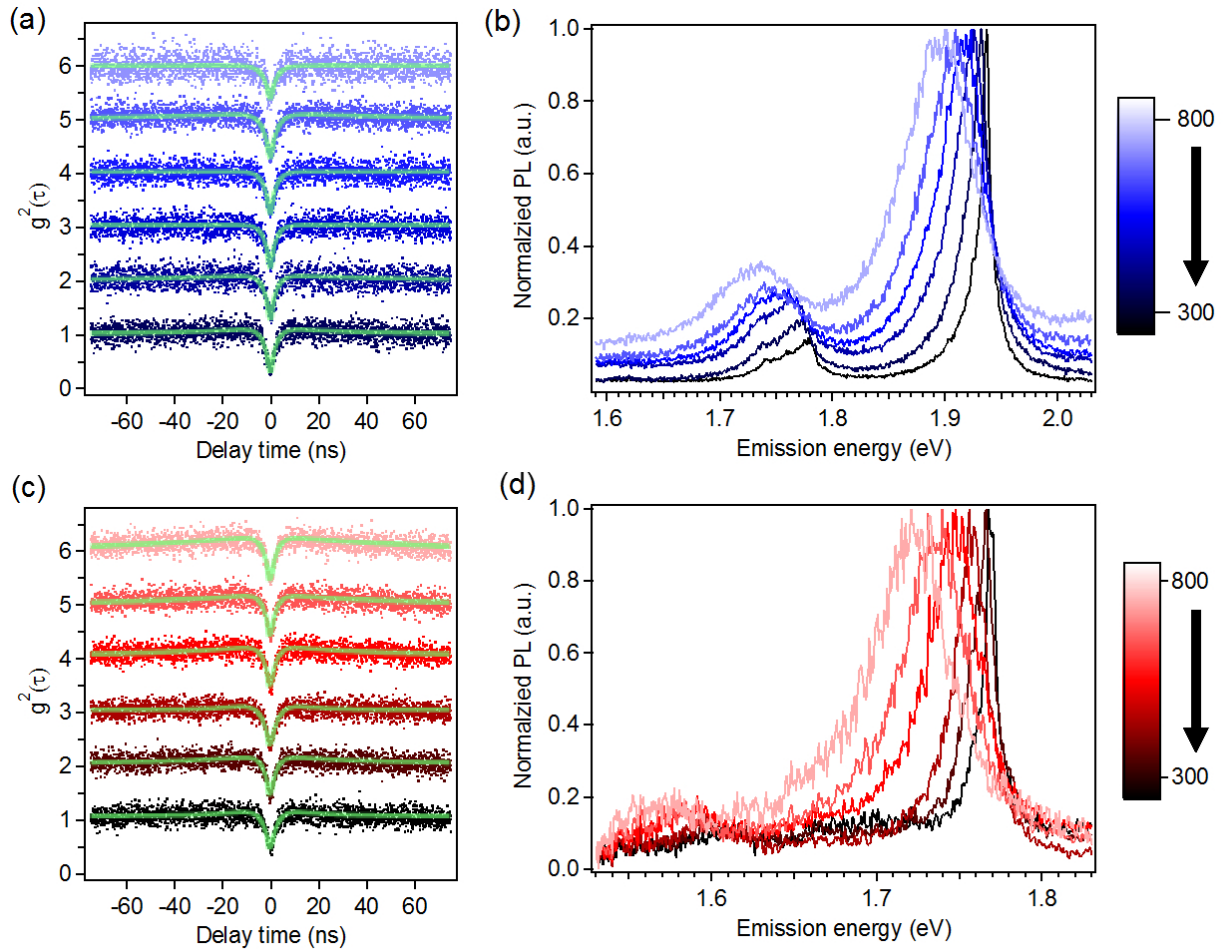

Figure 3. Temperature dependence of the quantum emission and photoluminescence of SPSs in hBN during cooling, (a) autocorrelation and (b) normalized PL spectrum measured at 100 K intervals from 800 to 300 K from SPS #1. (c) and (d) are equivalent data from SPS #2. The autocorrelation functions in (a) and (c) were offset for clarity. Solid curves are fit with a three level system.

To analyze the efficiency and purity of the sources, $g^2(0)$ is plotted as a function of temperature for SPS #1 and SPS #2 in figure 5a and d, respectively. The data were collected during both the heating (full symbols) and cooling ("+" and "×" symbols) phases of the thermal cycle for each SPS. The results show that the photon purity is constant over the entire temperature range, with $g^2(0) \approx 0.25$ (0.4) for SPS #1 (SPS #2). The background-corrected $g^2(0)$ for each source (light blue, and light orange) is also constant across the entire temperature range. To the best of our knowledge, these temperatures are the highest reported for a stable, operating quantum system, surpassing other materials systems, such as gallium nitride, by more than 400 K[27, 28]. We note that the scatter in the raw $g^2(0)$ values seen in figure 3a, d is caused by thermal drift (and hence drift in the background intensity) during data acquisition.

To further analyze the effects of temperature on the optical properties of both SPSs, we recorded the ZPL energy, linewidth, and intensity as a function of temperature. The ZPL position of each source red shifts by ~40 meV upon heating to 800 K (figure 5b, e). A red shift is expected and attributed primarily to expansion of the substrate lattice upon heating, and electron-lattice interactions[29]. The ZPL linewidth broadens at a linear rate of 0.13 meV K$^{-1}$ and 0.11 meV K$^{-1}$ for SPS #1 and SPS #2, respectively (figure 5b, e), which is caused primarily by interactions with lattice phonons[22]. The ZPL red-shift and broadening are reversible upon cooling back to room temperature, as shown by "×" and "+" symbols in figure 5b,e.

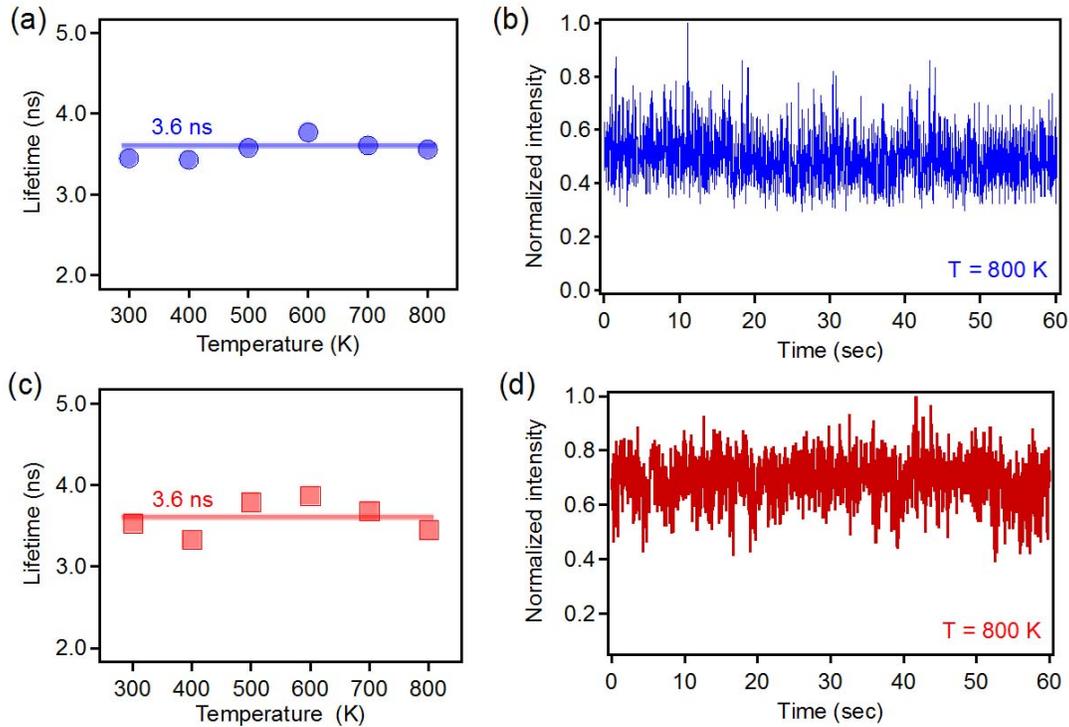

Figure 4. Fluorescence lifetime and emission stability of the SPSs in hBN. (a) Fluorescence lifetime of SPS #1 measured at 100 K increment between 300 K and 800 K. A 30 µW, 510 nm pulsed laser was used as the excitation source. (b) Normalized fluorescent of SPS #1 at 800 K for 60 seconds showing the photostablity of the source. (c) and (d) show the equivalent data for SPS #2.

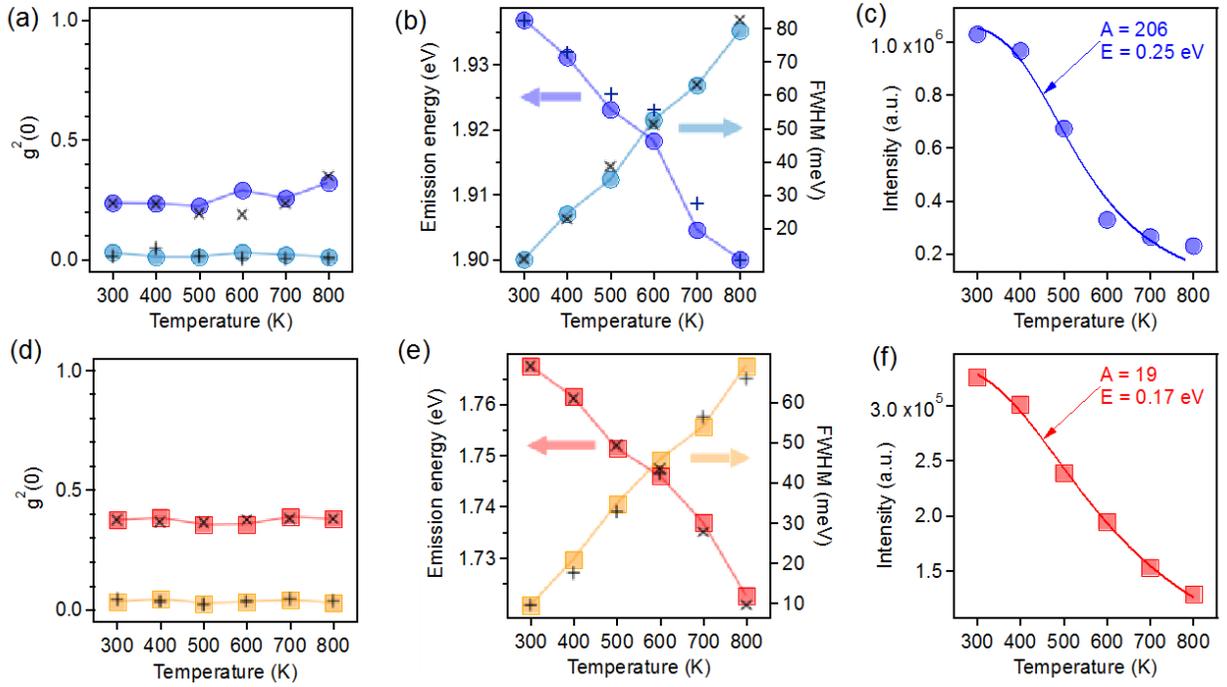

Figure 5. Photophysical properties of hBN single photon sources measured during the thermal cycle. (a) Coincidence counts at zero delay time, g²(0) (blue circles) and the background-corrected values (light blue circles) obtained for SPS #1 as a function of temperature during the heating phase of the thermal cycle. Corresponding values obtained during the cooling phase are shown as "×" and "+" symbols). (b) Zero phonon line (ZPL) position (left) and width (right) plotted as a function of temperature during heating and cooling. In (a-b) the solid lines are guides to the eye. (c) ZPL intensity versus temperature for SPS #1 (blue circles). The data are fit using equation 1 (solid curve). (d-f) Equivalent data for SPS #2.

The ZPL intensity decreases by a factor of ~ 5 and ~ 2 upon heating to 800 K for SPS #1 (figure 5c) and SPS #2 (figure 5f), respectively. A decrease is expected due to an increase in the efficiency of competing non-radiative recombination pathways with increasing temperature. Assuming one nonradiative pathway with activation energy E, the dependence of ZPL intensity (I) on temperature (T) can be described by[30]:

$$I = \frac{I_0}{(1+A\exp(-E/k_bT))} \quad (1)$$

Here, $I_0$ is the emission intensity at 0 K, A is a system-specific pre-exponential factor related to the nonradiative transition rate in the limit E→0, and $k_b$ is Boltzmann's constant. Figure 6 shows a four level system used to explain the temperature-dependence of SPS#1 and SPS#2 in the framework of equation 1.

The experimental data from each source can be fit well using equation 1 (solid lines in figure 5c, f) only if both E and A are varied. The activation barriers are high in both cases (E ~ 0.25 eV and ~0.17 eV for SPS #1 and SPS #2, respectively) as expected, since bright luminescence is observed from both sources even at 800 K. However, the barriers do not explain the difference between the quenching rates of the two SPSs seen in figure 5c, f which is instead caused by a large difference in the pre-factor A (~206 and ~19 for SPS#1 and SPS #2, respectively). A is related to the transition rate $k_{24}/k_{42}$ shown in figure 6.

To understand the difference in A between SPS #1 and SPS #2, we consider the Huang-Rhys (HR) factor (estimated from the relative intensity of the ZPL and the PSB[31, 32]) for each source. The values of the HR factors for SPS #1 (SPS #2) are 0.5 (0.3). The relatively low HR values indicate weak electron-phonon coupling in both cases, consistent with the fact that the sources operate efficiently even at 800 K. The fact that the HR factor of SPS #2 is smaller is consistent with the observation that it quenches more slowly upon heating (figure 5c, f), and is likely the reason for why A is smaller for SPS #2 than for SPS #1.

The difference is attributed to environmental fluctuations, which are believed to be responsible for the broad range of ZPL energies, widths and HR factors reported previously for quantum emitters in hBN[19, 22, 23].

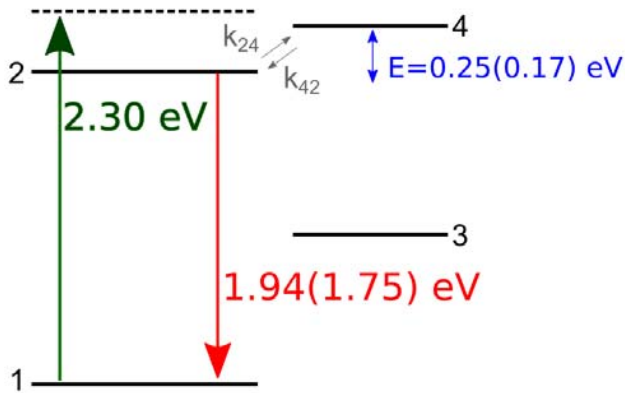

Figure 6. Simplified four level system used to explain the temperature-dependence of SPSs in hBN. The ground and excited states are labeled 1 and 2, 2.3 eV is the excitation energy used in our experiments, 1.94 eV (1.75 eV) is the ZPL energy of SPS #1 (SPS #2), E is the activation barrier in Equation 1, which equals 0.25 eV (and 0.17 eV) for SPS#1 (SPS#2), and $k_{24}$ and $k_{42}$ signify transition rates between states 2 and 4. The nonradiative recombination pathway proceeds through states 2, 4 and 3.

## CONCLUSIONS

In conclusion, our work demonstrates robust SPSs that can operate at elevated temperatures of up to 800 K. The results open up exciting new possibilities to integrate hBN SPSs with large scale, on-chip quantum photonic and optotelectronic devices that work under ambient conditions or elevated temperatures. Our results also provide vital information on stability, phonon coupling and photophysical properties of defects in layered materials, stimulating further research in this area.

## AUTHOR INFORMATION

**Corresponding Author**
\* Email: igor.aharonovich@uts.edu.au
^Email: milos.toth@uts.edu.au

**Author Contributions**
M.K., B. R. and T.T.T. contributed equally to this work. M.K. and B.R. carried out the experiments and M.K. and T.T.T. analyzed the data. The manuscript was prepared by contribution of all authors. I.A. and M.T. supervised the project.

## ACKNOWLEDGMENT

Financial support from the Australian Research Council (via DP140102721, IH150100028, DE130100592), FEI Company, and the Asian Office of Aerospace Research and Development grant FA2386-15-1-4044 are gratefully acknowledged.

## REFERENCES

1. Yuan Z, Kardynal BE, Stevenson RM, Shields AJ, Lobo CJ, Cooper K, *et al.* Electrically Driven Single-Photon Source. *Science* 2002, **295**(5552)**:** 102.

2. Mizuochi N, Makino T, Kato H, Takeuchi D, Ogura M, Okushi H, *et al.* Electrically driven single-photon source at room temperature in diamond. *Nature Photonics* 2012, **6**(5)**:** 299-303.

3. Sun Y, Zhou K, Sun Q, Liu J, Feng M, Li Z, *et al.* Room-temperature continuous-wave electrically injected InGaN-based laser directly grown on Si. *Nature Photonics* 2016, **10**(9)**:** 595-599.

4. Mamin HJ, Kim M, Sherwood MH, Rettner CT, Ohno K, Awschalom DD, *et al.* Nanoscale Nuclear Magnetic Resonance with a Nitrogen-Vacancy Spin Sensor. *Science* 2013, **339**(6119)**:** 557-560.

5. Christle DJ, Falk AL, Andrich P, Klimov PV, Hassan JU, Son NT, *et al.* Isolated electron spins in silicon carbide with millisecond coherence times. *Nat Mater* 2015, **14**(2)**:** 160-163.

6. Kuzmich A, Bowen WP, Boozer AD, Boca A, Chou CW, Duan LM, *et al.* Generation of nonclassical photon pairs for scalable quantum communication with atomic ensembles. *Nature* 2003, **423**(6941)**:** 731-734.

7. O'Brien JL, Furusawa A, Vučković J. Photonic quantum technologies. *Nature Photonics* 2009, **3**(12)**:** 687-695.

8. Awschalom DD, Bassett LC, Dzurak AS, Hu EL, Petta JR. Quantum Spintronics: Engineering and Manipulating Atom-Like Spins in Semiconductors. *Science* 2013, **339**(6124)**:** 1174.

9. Ladd TD, Jelezko F, Laflamme R, Nakamura Y, Monroe C, O'Brien JL. Quantum computers. *Nature* 2010, **464**(7285)**:** 45-53.

10. Aharonovich I, Englund D, Toth M. Solid-state single-photon emitters. *Nat Photon* 2016, **10**(10)**:** 631-641.

11. Fedorych O, Kruse C, Ruban A, Hommel D, Bacher G, Kümmell T. Room temperature single photon emission from an epitaxially grown quantum dot. *Applied Physics Letters* 2012, **100**(6)**:** 061114.

12. Kako S, Santori C, Hoshino K, Gotzinger S, Yamamoto Y, Arakawa Y. A gallium nitride single-photon source operating at 200 K. *Nat Mater* 2006, **5**(11)**:** 887-892.

13. Castelletto S, Johnson BC, Ivády V, Stavrias N, Umeda T, Gali A*, et al.* A silicon carbide room-temperature single-photon source. *Nat Mater* 2014, **13**(2)**:** 151-156.

14. Kurtsiefer C, Mayer S, Zarda P, Weinfurter H. Stable Solid-State Source of Single Photons. *Physical Review Letters* 2000, **85**(2)**:** 290-293.

15. Aharonovich I, Castelletto S, Simpson DA, Su CH, Greentree AD, Prawer S. Diamond-based single-photon emitters. *Reports on Progress in Physics* 2011, **74**(7)**:** 076501.

16. Cassabois G, Valvin P, Gil B. Hexagonal boron nitride is an indirect bandgap semiconductor. *nature Photonics* 2016, **10:** 262-266.

17. Lin Y, Connell JW. Advances in 2D boron nitride nanostructures: nanosheets, nanoribbons, nanomeshes, and hybrids with graphene. *Nanoscale* 2012, **4**(22)**:** 6908-6939.

18. Liu Z, Gong Y, Zhou W, Ma L, Yu J, Idrobo JC*, et al.* Ultrathin high-temperature oxidation-resistant coatings of hexagonal boron nitride. *Nat Commun* 2013, **4:** 2541.

19. Tran TT, Elbadawi C, Totonjian D, Lobo CJ, Grosso G, Moon H*, et al.* Robust Multicolor Single Photon Emission from Point Defects in Hexagonal Boron Nitride. *ACS Nano* 2016, **10**(8)**:** 7331-7338.

20. Bourrellier R, Meuret S, Tararan A, Stéphan O, Kociak M, Tizei LHG*, et al.* Bright UV Single Photon Emission at Point Defects in h-BN. *Nano Letters* 2016, **16**(7)**:** 4317-4321.

21. Tran TT, Zachreson C, Berhane AM, Bray K, Sandstrom RG, Li LH*, et al.* Quantum Emission from Defects in Single-Crystalline Hexagonal Boron Nitride. *Physical Review Applied* 2016, **5**(3)**:** 034005.

22. Jungwirth NR, Calderon B, Ji Y, Spencer MG, Flatte ME, Fuchs GD. Temperature Dependence of Wavelength Selectable Zero-Phonon Emission from Single Defects in Hexagonal Boron Nitride. *Nano Lett* 2016.

23. Chejanovsky N, Rezai M, Paolucci F, Kim Y, Rendler T, Rouabeh W, *et al.* Topological attributes and photo-dynamics of visible spectrum quantum emitters in hexagonal boron nitride. *arXiv:160803114* 2016.

24. Martínez LJ, Pelini T, Waselowski V, Maze JR, Gil B, Cassabois G, *et al.* Efficient single photon emission from a high-purity hexagonal boron nitride crystal. *Physical Review B* 2016, **94**(12): 121405.

25. Choi S, Tran TT, Elbadawi C, Lobo CJ, Wang X, Juodkazis S, *et al.* Engineering and localization of quantum emitters in large hexagonal boron nitride layers. *arXiv:160804108* 2016.

26. Beveratos A, Brouri R, Gacoin T, Poizat J-P, Grangier P. Nonclassical radiation from diamond nanocrystals. *Physical Review A* 2001, **64**(6).

27. Holmes MJ, Choi K, Kako S, Arita M, Arakawa Y. Room-temperature triggered single photon emission from a III-nitride site-controlled nanowire quantum dot. *Nano Lett* 2014, **14**(2): 982-986.

28. Holmes MJ, Kako S, Choi K, Arita M, Arakawa Y. Single Photons from a Hot Solid-State Emitter at 350 K. *ACS Photonics* 2016, **3**(4): 543-546.

29. Varshni YP. Temperature dependence of the energy gap in semiconductors. *Physica* 1967, **34**(1): 149-154.

30. Lagomarsino S, Gorelli F, Santoro M, Fabbri N, Hajeb A, Sciortino S, *et al.* Robust luminescence of the silicon-vacancy center in diamond at high temperatures. *AIP Advances* 2015, **5**(12): 127117.

31. Huang K, Rhys A. Theory of Light Absorption and Non-Radiative Transitions in F-Centres. *Proceedings of the Royal Society of London Series A Mathematical and Physical Sciences* 1950, **204**(1078): 406.

32. Neu E, Steinmetz D, Riedrich-Moeller J, Gsell S, Fischer M, Schreck M, *et al.* Single photon emission from silicon-vacancy centres in CVD-nano-diamonds on iridium *New Journal of Physics* 2011, **13**: 025012.